\RequirePackage{fix-cm}
\documentclass[smallextended]{svjour3}       
\smartqed  
\usepackage{graphicx}
\usepackage[ruled,vlined]{algorithm2e}
\usepackage{algpseudocode}
\usepackage{LaTeX_Supporting_File_orcidlink}
\journalname{Computational Statistics}

\begin{document}

\title{Study of inter-individual variability of three-dimensional data table}
\subtitle{Detection of unstable variables and samples}


\author{Lo\"ic Labache \and
        Marc Joliot \and
        Ga\"elle E.~Doucet \and
        J\'er\^ome Saracco
}


\institute{Lo\"ic Labache\, \orcidlink{0000-0002-5733-0743}   \and Marc Joliot\, \orcidlink{0000-0001-7792-308X} 
              \at Groupe d'Imagerie Neurofonctionnelle – CEA {\rm \&} IMN, UMR 5293 – Centre Broca Nouvelle-Aquitaine, 146 rue L\'eo Saignat, 33000 Bordeaux, France \\
              \email{loic.labache@u-bordeaux.fr} 
              \email{marc.joliot@u-bordeaux.fr} 
            \and
           Gaelle E. Doucet\, \orcidlink{0000-0003-4120-0474} 
              \at Boys Town National Research Hospital, Omaha, NE, USA \\
              \email{gaelle.doucet@boystown.org} 
            \and
           J\'er\^ome Saracco\, \orcidlink{0000-0003-4198-4002} \and Lo\"ic Labache 
              \at Inria Bordeaux Sud Ouest {\rm \&} Institut de Math\'ematiques de Bordeaux, UMR CNRS 5251 {\rm \&} ENSC Bordeaux INP, 109 Avenue Roul, 33400 Talence, France \\
              \email{jerome.saracco@math.u-bordeaux.fr}
}

\date{First submitted on 26 Mars 2019}

\maketitle

\begin{abstract}
We propose two methodologies in order to better understand the inter-individual variability of resting-state functional Magnetic Resonance Imaging (fMRI) brain data. The aim of the study was to quantify whether the average dendrogram is representative of the initial population and to identify its possible sources of instability. The average dendrogram is based on the Pearson correlation between resting-state networks. The first method identifies networks that can lead to unstable partitions of the average dendrogram. The second method identified homogeneous sub-samples of participants for whom their associated average dendrograms were more stable than that of the whole sample. The two suggested methods have shown significant quantifiable behavioral data results with regards to detecting an unstable network or presence of subpopulations when the noise level does not conceal the structure of the data. These two methods have been successfully applied to establish a cerebral atlas for late adulthood. The first method made it clear that there was no unstable network among the atlas networks. The second method highlighted the presence of two distinct sub-populations with different age-related brain organizations. 
\end{abstract}

\section{Introduction}
\label{intro}
The analysis of the partitions stability of a dendrogram is a crucial issue in order to check the replicability of the selected partitions. 

For classical two-dimensional data, it is possible to obtain a measure of the stability of the clusters (obtained from a given partition) using the approximately unbiased p-value (AU-value) obtained with a multiscale bootstrap resampling method (MBR: \cite{RefA}) available via the {\tt pvclust} function of the R library of the same name \cite{RefB}. The AU-value then indicates the unbiased frequency of occurrence of a cluster across the reference population of the dendrogram.

There is currently no counterpart to this method for three-dimensional data. This paper thus develops a similar method to {\tt pvclust} that is adapted to any kind of three-dimensional data and, in particular, for resting-state functional Magnetic Resonance Imaging (rs-fMRI) data.

fMRI allows to obtain regional brain electrical activity (\cite{RefC}, \cite{RefD}) by monitoring relative fluctuations in the Blood Oxygen Level Dependent (BOLD) signal.

The study of human brains' resting state functional organization consists in analyzing the synchronicity between the BOLD signals of different brain areas or networks. Commonly, the synchronicity study is performed using the calculation of Pearson correlation coefficients between the BOLD signals of all network pairs, resulting in correlation matrices of dimension $K\times K$ for $K$ brain networks. There are databases of several hundred individuals containing 3D structures (of dimension $K\times K$) that gather all of the correlation matrices $M_s,~s=1,..., S$ of the $S$ individuals in the database.

For a some databases, in this case, the BIL\&GIN database, \cite{RefE}), the first work consisted of a agglomerative hierarchical clustering of $K$ brain networks. The second step was to find an optimal number of clusters reflecting the best resting state cerebral organisation \cite{RefF}. The methodology consisted of averaging all the Fisher transformations of the matrices $M$ in the following way:

\begin{equation}
M_m=\tanh \left( \frac{\sum_{s=1}^S\mbox{arctanh}(M_s)}{S}\right)
\end{equation}

This matrix is then transformed into a dissimilarity matrix $D_m$~: $D_m=(1-M_m)/2.$. Then $m$ the aggregation of the $K$ brain networks is accomplished via agglomerative hierarchical clustering(according to Ward’ method)based on $D_m$.

Taking into account the inter-individual variability, \cite{RefF} adapted a procedure called {\tt pvclust} \cite{RefG} which allows researchers to assess the uncertainty associated with the different partitions of the hierarchical clustering through a p-value obtained by bootstrapping. The previous methodology provides a Bootstrap Probability p-value (called BP-value) and after correction, an Approximately Unbiased p-value (denoted AU-value). The AU-values are calculated using the bootstrap values recommended in \cite{RefB} by using anywhere from 50\% to 140\% of the sample. These p-values indicate how well the different partitions are supported by the data.

\begin{figure}[htbp!]
\begin{center}
  \includegraphics[width=1.00\textwidth]{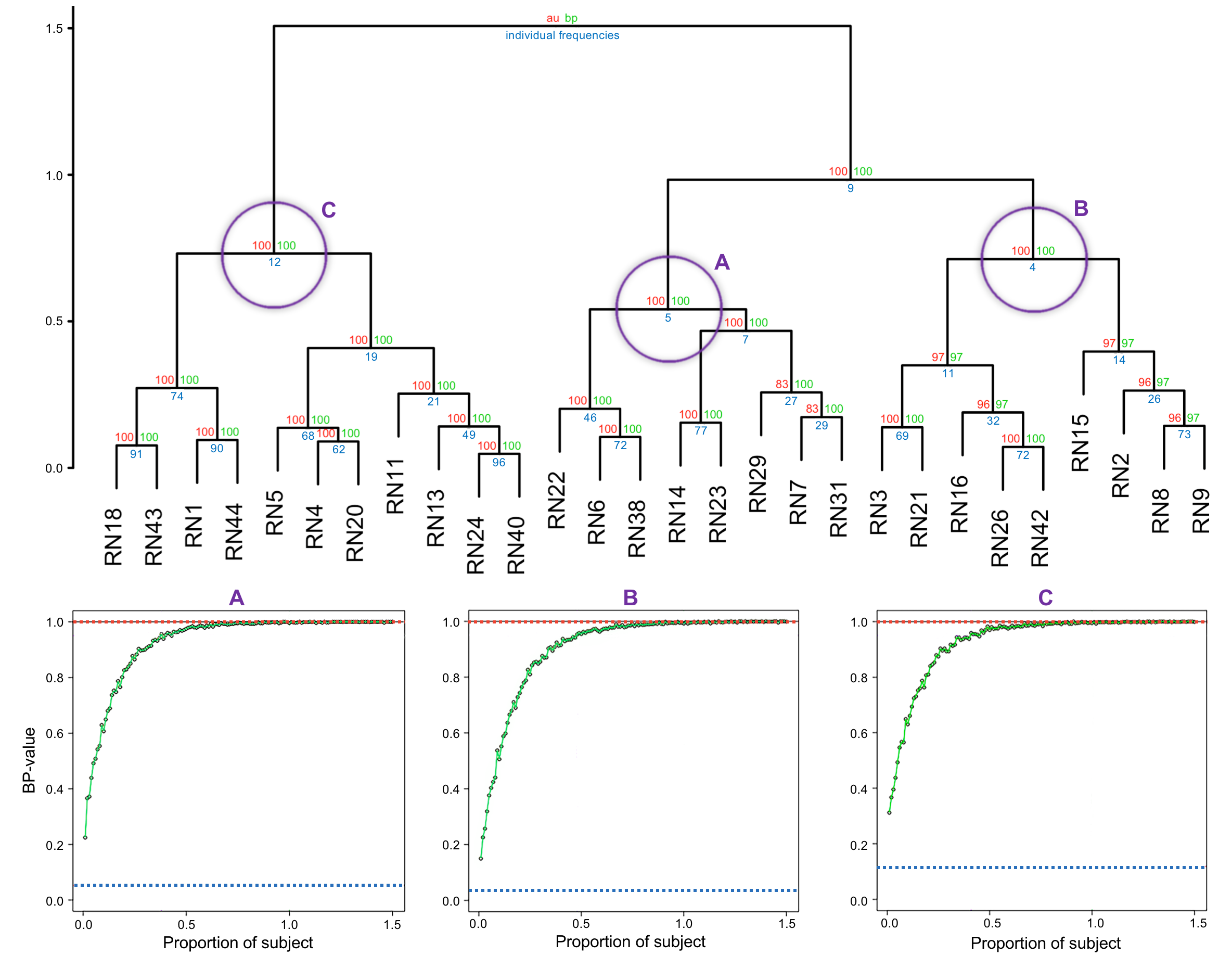}
\caption{Agglomerative hierarchical clustering from $M_m$ (top). Evolution of BP-value as a function of the proportion of individuals sampled in the bootstrap (below). The individual frequency of the clusters are in blue. The BP-values are in green. The AU-values are in red.}
\label{fig:1}
\end{center}
\end{figure}

To illustrate this methodology, let us consider a population of 439 subjects. The optimal number of clusters was determined using the R library {\tt NbClust} \cite{RefH}. This package provides 30 statistical indices for determining the optimal number of clusters and offers the best clustering scheme from the different results obtained by varying all combinations of the number of clusters for the chosen method, in this case, hierarchical clustering with Ward’s criterion. We selected the number of clusters that satisfied a maximum of indices and found it to be equal to 3. For the corresponding three optimal clusters the associated BP and AU-values are equal to 100\%  (Fig.~\ref{fig:1}). \cite{RefB} recommends using AU-values, these 3 retained clusters were perfectly represented by our data. However, the AU-values represent the stability of the $M_m$ based clusters and not on the empirical frequency of appearance through these clusters across the $S$ subjects, called individual frequencies below. Fig.~\ref{fig:1} illustrates this event. We can see the evolution of the BP-values as a function of the proportion of subjects used in the bootstrap sampling, as well as the actual value of the individual frequencies of cluster occurrence. For example, for the considered partition of 3 clusters, the corresponding three BP-values are 100\%,  whereas the individual frequencies of appearance are respectively 5\% , 4\%  and 12\%  for these 3 clusters. The individual frequencies of occurrence correspond to the number of times a cluster appears among the individual dendrograms of the population. In the study of intrinsic brain organization, these proportions are unreliable and, in a general framework, this shows that the {\tt pvclust} algorithm cannot be used for three-dimensional data; the BP-values do not reflect the individual frequencies.

In this paper, two methodologies are proposed to better understand the inter-individual variability of partitions from $M_m$ : 
\begin{itemize}
    \item the first one allows the identification of networks that can lead to unstable partitions from $M_m$;
    \item the second one allows the identification of homogeneous subpopulations of subjects  across the population by maximizing the individual frequency of appearance of clusters from the partitions built on their associated $M_m$ matrices.
\end{itemize}

In the first part of this paper, we will present how to calculate the empirical frequency $Q_k$ through a network $R_k$ based on the comparison of its position in the average dendrogram with respect to the individual dendrograms. We will present how to use it to identify an unstable network at the population level. Finally, from the individual frequency of a cluster, we show how to extract, from the initial population, subpopulations that are, on average, more stable.

\section{Presentation of the proposed approaches}
\label{sec:2}
Consider  $M_s,~s=1,..., S$, the $S$ individual correlation matrices of dimension $K\times K$.

Starting from the hierarchical ascendant clustering based on the mean matrix $M_m$ defined in the introduction, $K$ nested partitions$P_m^j,~j=1,\dots,K$ can be obtained and summarized by vectors of size $K$ defined as follows: for the partition (also called pattern afterwards) into $j$ clusters, the $k$\textsuperscript{\textit{th}} component of $P_m^j$ is equal to 1 if the network $R_k$ appears in the new cluster created by switching from a partition into $(j+1)$ clusters to a partition into $j$ clusters . In the following, we will note:

\begin{equation}
P_m=\{P_m^1,\dots,P_m^j,\dots,P_m^K\}
\end{equation}

with $P_m^1=(1,\dots,1)$ and $P_m^K=(0,\dots,0)$ by convention.

In the same way, for each subject $s$, the corresponding $K$ nested partitions can be constructed from the $M_s$ matrices:

\begin{equation}
P_s=\{P_s^1,\dots,P_s^j,\dots,P_s^K\}~~\mbox{for}~s=1,\dots,S
\end{equation}

For each non-trivial1e\footnote{Partitioning into one cluster (respectively into $K$ clusters ) is considered as trivial.} partition from $M_m$ , we will search, among the $(K-2)$ non-trivial partitions of each subject $s$, the element of $P_s$ closest to $P_m^j$ based on the Sorensen-Dice index defined as follows:

\begin{equation}
C_s^j=1-\min_{l=2,\dots,K-1}\frac{2(P_m^j)'P_s^l}{(P_m^j)'P_m^j+(P_s^l)'P_s^l}
\end{equation}
where the notation $v'$ denotes the transpose of the vector $v$. A zero index $C_s^j$ (respectively equal to 1) indicates that the vectors $P_m^j$ and $P_s^j$ are equal (respectively, do not share a common network).

In the following, we call \textit{alternative pattern} to $P_m^j$, denoted by $\tilde{P}^j$ hereafter, the pattern(s) of $P_s$ closest to $P_m^j$ in the sense of the index $C_s^j$ (if this pattern is not equal to the average pattern $P_m^j$).

From the set of the $\tilde{P}^j$, it is then possible to calculate a $Q_k$ score of participations for each network $R_k$ . This score $Q_k$ represents the empirical frequency (expressed in \%) that the network $R_k$ is constituted of an alternative pattern, whatever the average pattern. For an average pattern $P_m^j$, let us define $F^j$ as the number of single subjects expressing the alternative pattern $\tilde{P}^j$, and let $F_k^j$ be the number of single subjects expressing the alternative pattern $\tilde{P}^j$ containing the network $k$. The empirical frequency $Q_k$ of a network $R_k$ is then defined as follows:
 
\begin{equation}
Q_k=\frac{\sum_{j=1}^{K-2}F_k^j}{\sum_{j=1}^{K-2}F^j}
\end{equation}
Note that an alternative pattern of $P_m^j$ can never be equal to an average pattern $P_m^j$ of another level than $l\neq j$.

\subsection{Method for identifying an unstable network}
\label{sec:2_1}
From the participation score $Q_k$ of each network $R_k$, it is possible to identify the one that will be said to be \textit{unstable} among the $K$ networks (i.e. a network that will not be stable with respect to Pm across the subjects). Typically, a network $R_{k^\ast}$ that “roams” on the dendrogram associated with $P_m$ without having a fixed attachment within the $S$ individual dendrograms $P_s$ leads to an unstable mean tree. The network showing the most marked instability is obtained by : 

\begin{equation}
k^*=\arg\max_{k=1,\dots,K}Q_k
\end{equation}

By convention, if $Q_1=\dots=Q_K$, it is considered that there are no unstable networks. 

\subsection{Method for identifying subpopulations of subjects}
\label{sec:2_2}
Let us first define the individual frequencies of the \textit{average} dendrogram patterns $P_m^j$ denoted $F_m^j$ as the number of single subjects expressing $P_m^j$.

First, the methodology consists of iteratively selecting the subjects belonging to the pattern $P_m^{j^\ast}$ such that $j^*=\arg\min_{j=2,\dots,K-1}F_j$. At each iteration, the selected subjects then form a new subpopulation, and the new current population is composed of the unselected subjects. The stopping criterion is to obtain a tree in the current subpopulation such that $\forall j, F_m^j=1$.

In order to counteract the presence of noise in the data, the cophenetic correlation $d$ \cite{RefI} between each of the dendrograms obtained from the mean matrix of each identified subpopulation is then calculated. This allows to group subjects belonging to the same subpopulations based on cophenetic correlations above a fixed threshold $t$. The choice of $t$ is left to the user: however it should be noted that: 1) the threshold $t$  directly influences the creation of subpopulations and 2) it allows the reduction of noise in the tridimensional data, if noise is present. A high value of $t$ (i.e. close to $1$) will be recommended if similar populations in the three-dimensional data is suspected, close to $0.5$ otherwise. To do this, we set up the following iterative procedure. Initially, we calculate the matrix of cophenetic correlations between the mean dendrograms of the subpopulations obtained at the end of the first step. Thereafter, we threshold this matrix according to $t$ (i.e. correlations lower than $t$ are set to $0$). From this matrix, we construct the d-weighted graph G, where each node represents a subpopulation. Thanks to the Louvain method \cite{RefJ}, an optimal partition number is obtained by maximizing the modularity of the $G$ graph. Subsequently, the subpopulations are grouped together according to the Louvain partitioning, then a new cophenetic correlation matrix is computed from the mean dendrograms of these new subpopulations. Louvain partitioning procedure is then repeated until $G$ has only one node. 

This procedure (Algorithm~\ref{alg:1}) leads to iteratively extracting the most stable subjects from the \textit{average} current $P_m^{current}$ dendrogram. It should be noted that by construction, $P_m^{current}$ will vary in terms of the arrangement and composition of its scores.

Finally, this procedure is able to identify the existence of a subpopulation of subjects with a pattern that differs constantly within it, or identify the existence of a subpopulation that does not have a strong structure in terms of dendrogram. 

\medskip

\begin{algorithm}[H]
\label{alg:1}
\SetAlgoLined
\KwIn{$F_m^j, individuals, P_m^j$}
 \While{$\forall j, F_m^j = 1$}{
  ${group}_i \leftarrow individuals \in P_m^{j^\ast}, j^\ast = \arg\min_{j = 2, ..., K-1} F^j$\;
  $curentPopulation \leftarrow individuals \notin {group}_t$\;
  $(F_m^j , P_m^j , {Dend}_i) \leftarrow frequency\_extraction(curentPopulation)$\;
 }
  \While{$|graphFromMatriceD| == 1$}{
  $matriceD \leftarrow cophenetic({Dend}_i)$\;
  $matriceD < t \leftarrow 0$\;
  $modules \leftarrow Louvain(graphFromMatriceD)$\;
  $subPopulation \leftarrow individuals \in modules$\;
  }
  \KwOut{$subPopulation$}
 \caption{Identification of subpopulations}
\end{algorithm}

\medskip

\noindent
where \textit{frequency\_extraction} is the implemented algorithm described in Sect.~\ref{sec:2} in order to extract the individual frequencies of the average dendrogram partitions, \textit{cophenetic} is the function that computes the matrix of cophenetic correlations between all the dendrograms two by two of a list of dendrograms, \textit{Louvain} is the implemented Louvain algorithm for identifying modules from a weighted graph $G$ (here each node of the graph is a dendrogram composed of $s$ individuals), and \textit{dendrograms} is the function computing the dendrogram from $M_m$.

\section{Study of numerical behavior}
\label{sec:3}
The numerical behaviour of the two proposed methodologies is illustrated on simulated \textit{fMRI-type} data. 
\begin{itemize}
    \item In the first simulation study, we consider a population of $S$ subjects whose correlation matrices $M_s, s = 1, ..., S$ are \textit{close} (in terms of \textit{mean} dendrogram), except for a network $R^{k^\ast}$ which is randomly permuted for each subject. The objective here is to detect this unstable network which makes the \textit{mean} dendrogram $P_m$ not very representative of the population of $S$ subjects.
    \item In the second simulation study, the population of S subjects is composed of two subpopulations of homogeneous subjects (in terms of the \textit{average} dendrogram). The objective here is to identify these two subpopulations.
\end{itemize}

\paragraph{\textbf{Data generation.}} In order to create subjects \textit{close to each other} in the population (or subpopulation), we introduce a correlation matrix $M$ that we will \textit{noise} with a Gaussian random error $\varepsilon$ centered with a standard deviation $\sigma$ as follows: for $1 \leq i < l \leq K$,

\begin{equation}
M_{(i,l)}=\tanh(\mbox{arctanh}(M_{(i,l)})+e_{i,l})~~~\mbox{et}~~~M_{(l,i)}=M_{(i,l)}
\end{equation}

where $M_{(i,l)}$ denotes the element $(i,j)$ of the matrix $M$ and $e_{(i,l)}$ is a realization of the noise $\varepsilon$.
Different noise levels $\sigma$ were considered in the simulation study. Naturally, the higher the noise level, the less homogeneous the population (or subpopulation) generated. To make the simulation study interesting, we consider $\sigma$ values such that the \textit{mean} dendrogram is similar to the un-noised \textit{mean} dendrogram. 

Thereafter, we will work with data similar to those presented in the introduction (i.e. correlation matrices consisting of $K = 28$ resting-state networks $R_k$).

\subsection{Simulation $1$: identification of the unstable network}
\label{sec:3_1}
In this first simulation, we will generate for a noise level of $0 \leq \sigma \leq 1$ with a step of $0.1$, with $K$ groups of $S = 500$ subjects from one of the individual $M$ matrices presented in the introduction. For each noise level and each network, we have randomly permuted\footnote{i.e. switch the values of the corresponding row/column in the correlation matrix} a network $R_k$, i.e: for a given noise level $\sigma$, we have $K = 28$ groups of $S=500$ subjects. The objective is to detect, for each noise level and for each permutation, the targeted network $R_k$ (i.e. the permuted network) as the unstable network using the proposed methodology. 

\begin{figure}[htbp!]
\begin{center}
  \includegraphics[width=1.00\textwidth]{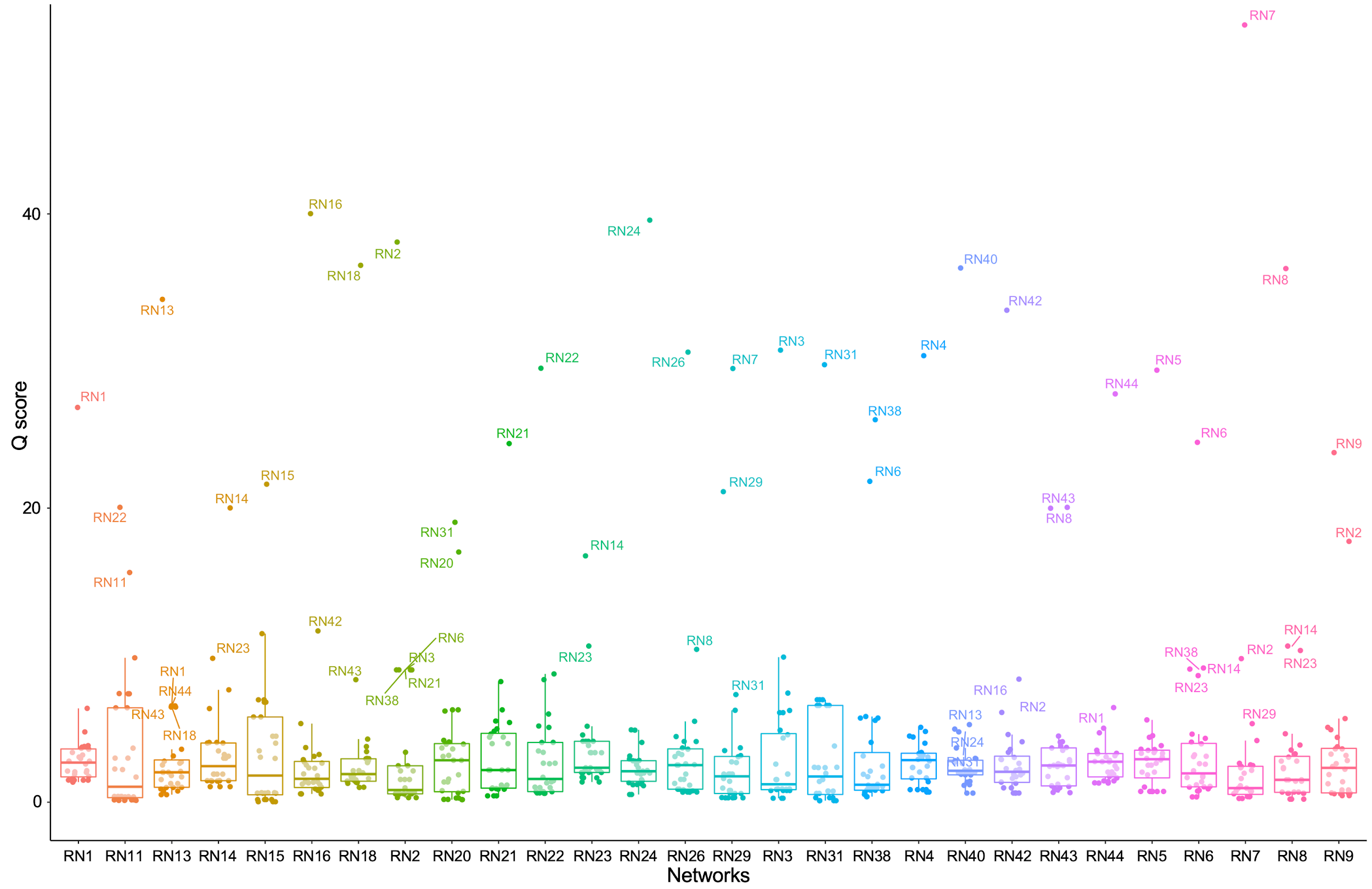}
\caption{Boxplot of the $Q_k$ score at $\sigma = 0$ for each permutation of $K$.}
\label{fig:2}
\end{center}
\end{figure}

\begin{figure}[htbp!]
\begin{center}
  \includegraphics[width=1.00\textwidth]{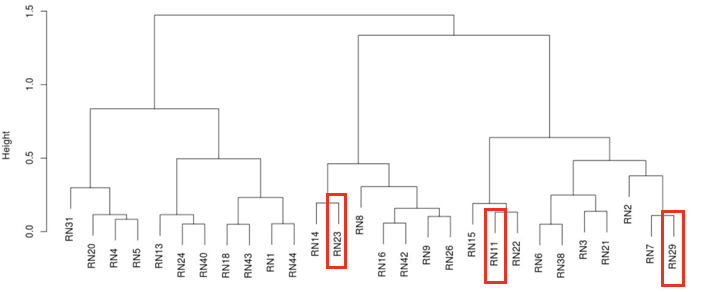}
\caption{\textit{Average} dendrogram (constructed from $M_m$) of the population of $S = 500$ subjects before the permutation of any network $R_k$. In the red box, permuted $R_k$ networks not detected as $k\ast$ at noise level $\sigma = 0$.}
\label{fig:3}
\end{center}
\end{figure}

Using the methodology proposed for a noise level $\sigma = 0$, we can see in Fig.~\ref{fig:2} that we manage in $89.3\%$ of the cases to detect the permuted network as the one with the highest $Q$ score. 
Looking more closely, we can see on the average dendrogram of the group before permutation (Fig.~\ref{fig:3}), that for the permutations of the networks $R_k$ ({\tt RN11}, {\tt RN23} and {\tt RN29}), the network $R^{k\ast}$ is detected each time the one which is directly linked to $R_k$ ({\tt RN22}, {\tt RN14} and {\tt RN7} respectively) instead of $R_k$ itself.

\begin{figure}[htbp!]
\begin{center}
  \includegraphics[width=0.95\textwidth]{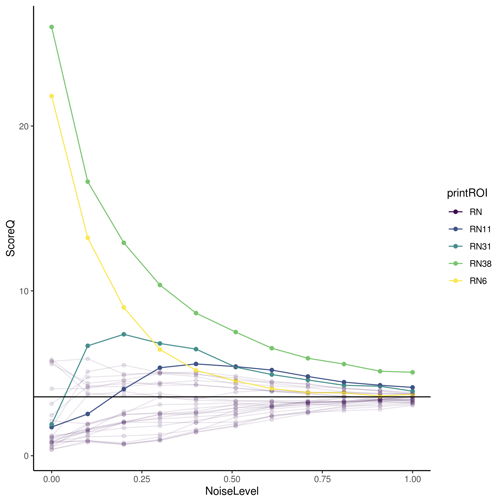}
\caption{Example of the evolution of the $Q_k$ score of the permuted network $R_k = {\tt RN38}$ as a function of the noise level $\sigma$ in a population of $S = 500$ subjects. The colored curves correspond to the 4 $R_k$ being most often detected as $R_{k^*}$. In green the target network {\tt RN38}, in yellow the network {\tt RN6} directly linked to {\tt RN38} in the initial dendrogram (see Fig.~\ref{fig:3})}
\label{fig:4}
\end{center}
\end{figure}

Taking the {\tt RN38} network as an example, when we look at the power of the method across all the different noise levels $\sigma$ (Fig.~\ref{fig:4}), we can see that we are able to detect the correct permuted network (i.e. $R_{k^*} = {\tt RN38}$) regardless of the noise level applied to the population of $500$ subjects whose network has been permuted. In this example, although RN38 (in green in Fig.~\ref{fig:4}) is always the network with the highest $Q$ score, the network {\tt RN6} (in yellow in Fig.~\ref{fig:4}) is very close behind. By removing the {\tt RN38} network from the whole population, we can see that the new mean dendrogram is much more stable with only a few alternative \textit{non-dominant} patterns (due to the different noise level); leading to a much higher individual frequency of the mean $P_m^k$ patterns, which is a very low standard deviation of the $Q_k$ score without a $k^\ast$ ever actually being detected. 

Globally, for 11 different noise levels from $0$ to $1$ (step of $0.1$), out of the $28$ possible permutations, we detect the correct $k\ast$ (i.e. the permuted network $R_k$) $95.1\%$ of the time. The $4.9\%$ error corresponds each time to the identification of the network directly linked to the network $R_k$ initially swapped for the simulation.

Our methodology allows us to detect the permuted $R_{k^*}$ network leading to the instability of the mean dendrogram and those regardless of the noise level injected in the data. Note that when $\sigma$ increases, (there is no more information / common structure in the data), all $Q_k$ scores will converge to $1/k$, i.e. each network participates in many alternative patterns or none of the networks participate in alternative patterns.

\subsection{Simulation $2$: identification of homogeneous subpopulations}
\label{sec:3_2}
\subsubsection{Detection of equidistributed population}
\label{sec:3_2_1}
In this part, we will consider two populations ($G_{50}^1$ and $G_{50}^2$) of $S = 500$ subjects. In each population, the subjects are equally divided into two subpopulations $A$ and $B$ of $250$ subjects each. For $G_{50}^1$, the proximity between the mean dendrogram of $A$ and $B$ is measured by the cophenetic correlation is $0.18$ (i.e. subpopulation $A$ is very distant from subpopulation $B$ in terms of the brain organization of their resting-state networks $R_k$). For $G_{50}^2$, the cophenetic correlation between $A$ and $B$ is $0.95$, which means they are two subpopulations whose brain organization is very close to each other.

Finally, for $G_{50}^1$ and $G_{50}^2$, we will consider different cases with a noise level $\sigma$ ranging from $0$ to $0.5$.

These two case studies, represented by $G_{50}^1$ and $G_{50}^2$, will allow us to evaluate the subpopulation detection power of our algorithm in addition to the presence of noise. Concerning the grouping threshold for the part of the method using the Louvain method; $t = 0.8$ for $G_{50}^1$ and $t = 0.99$ for $G_{50}^2$. Note that the threshold is more restrictive for $G_{50}^2$ since $A$ and $B$ are more similar in this population. 

The results for the $G_{50}^1$ study are provided in Tab.~\ref{tab:1}. For a null noise level, the algorithm perfectly identifies individuals as belonging to subpopulation $A$ or $B$. For a noise level $\sigma = 0.12$, the algorithm is able to identify $44\%$ of the individuals belonging to subpopulation $A$ and $36\%$ belonging to population $B$. For a noise level $\sigma = 0.25$, the algorithm finds more than $50\%$ of the individuals belonging to subpopulation $A$ ($28\%$ of individuals identified belonging to this subpopulation), but less than $25\%$ of the individuals belonging to subpopulation $B$. At the maximum noise level $\sigma = 0.50$, the algorithm finds $11\%$ of individuals from subpopulation $A$ and none from subpopulation $B$. 
\begin{table*}[]
\caption{Evolution of the detection of subpopulations $A$ and $B$ according to the evolution of the noise level $\sigma$ in Group $1$ ($G1$) and $2$ ($G2$). Equidistributed population: Group $G_{50}^1$,  cophenetic correlation between $A$ and $B$ of $0.18$, $50\%$ of subjects of each subpopulation $A$ and $B$. Group $G_{50}^2$,  cophenetic correlation between $A$ and $B$ of $0.95$, $50\%$ of subjects of each subpopulation $A$ and $B$. Not equidistributed population: Group $G_{75}^1$,  cophenetic correlation between $A$ and $B$ of $0.18$, $75\%$ of subjects of the subpopulation $A$ and $25\%$ of B. Group $G_{75}^2$,  cophenetic correlation between $A$ and $B$ of $0.95$, $75\%$ of subjects of the subpopulation $A$ and $25\%$ of $B$.}
\label{tab:1}   
\centering
\begin{tabular}{lllllllll}
\hline
\multicolumn{9}{l}{Equitably distributed population ($50\%$ $A$ \& $50\%$ $B$)}      \\
\hline\noalign{\smallskip}
 & \multicolumn{4}{l}{Subpopulation $A$} & \multicolumn{4}{l}{Subpopulation $B$}    \\
\hline\noalign{\smallskip}
                  & $\sigma = 0$ & $\sigma = 0.12$ & $\sigma = 0.25$ & $\sigma = 0.50$ & $\sigma = 0$ & $\sigma = 0.12$ & $\sigma = 0.25$ & $\sigma = 0.50$ \\
                  \hline\noalign{\smallskip}
Group 1           & 50    & 44       & 28       & 11       & 50    & 36       & 12       & 0        \\
Group 2           & 50    & 24       & 1        & 0        & 50    & 17       & 8        & 11       \\
\hline\noalign{\smallskip}
\multicolumn{9}{l}{Not equitably distributed population ($75\%$ $A$ \&   $25\%$ $B$)}                       \\
\hline\noalign{\smallskip}
 & \multicolumn{4}{l}{Subpopulation $A$}    & \multicolumn{4}{l}{Subpopulation $B$}    \\
\hline\noalign{\smallskip}
                  & $\sigma = 0$ & $\sigma = 0.12$ & $\sigma = 0.25$ & $\sigma = 0.50$ & $\sigma = 0$ & $\sigma = 0.12$ & $\sigma = 0.25$ & $\sigma = 0.50$ \\
                  \hline\noalign{\smallskip}
Group 1           & 75    & 65       & 64       & 0        & 25    & 11       & 0        & 0        \\
Group 2          & 75    & 58       & 33       & 10       & 25    & 0        & 21       & 0   \\
\hline
\end{tabular}
\end{table*}

The results for the $G_{50}^2$ study are shown in Tab.~\ref{tab:1}. In contrast to $G^1$, the mean dendrograms of subpopulations $A$ and $B$ are here similarly closer to each other according to their cophenetic distance (equal to $0.95$ for $G^2$, $0.18$ for $G^1$). This has a direct impact on the algorithm's ability to identify individuals as belonging to subpopulation $A$ or $B$. Concerning subpopulation $A$, the algorithm identifies correctly the individuals for a zero noise level, but the identification drops to $24$, $1$ and $0\%$ as it increases, which is worse than for $G^1$. There were the same results for the identification of subpopulation $B$, with results half as good as for $G^1$.

For a noise level $\sigma$ multiplied by $2$, the proportion of individuals identified as belonging to subpopulation $A$ or $B$ tends to be divided by $2$. Moreover, in view of the results, subpopulation $B$ of $G_{50}^1$ seems less stable on average than subpopulation $A$; the opposite is true for $G_{50}^2$.

\subsubsection{Detection of not equidistributed population}
\label{sec:3_2_2}
In this part we will consider two populations ($G^{1bis}$ and $G^{2bis}$) of $S = 500$ subjects. In each of the populations the subjects are equally divided into two subpopulations $A$ and $B$; of $375$ subjects for $A$ and $125$ for $B$. For $G^1$, the proximity between the mean dendrogram of $A$ and $B$ is the same as in the previous part, as the grouping threshold for the part of the method using the Louvain method.

Finally, for $G^{1bis}$ and $G^{2bis}$, we simulated different cases with a noise level $\sigma$ ranging from $0$ to $0.5$.

The simulation results are available in Tab.~\ref{tab:1}. Concerning subpopulation $A$ for $G^{1bis}$ and $G^{2bis}$, the algorithm identifies more than $85\%$ of its individuals up to a noise level $\sigma = 0.25$. As for the equitably distributed subpopulations, the algorithm manages to identify $100\%$ of the individuals belonging to the different subpopulations when the noise level is zero, and $0\%$ when $\sigma = 0.50$ (there is an exception for $G^{2bis}$ where $10\%$ of the individuals of subpopulation $A$ are correctly identified).

Again, we can note that the more noise increases one group disappears, either the $A$ group or the $B$ group. The more the number of subjects belonging neither to $A$ nor to $B$ increases up to a rate of $100\%$.

We alos note from  Tab.~\ref{tab:1} that: first the subjects detected as belonging to group $A$ or $B$ are initially subjects belonging to these groups (very low noise error rate). Secondly, it is interesting to note that the category of subjects neither belonging to group $A$ or $B$ is composed of similar proportions to the one we initially injected during the creation of the simulation.

\section{A real case study}
\label{sec:4}
\begin{figure}[htbp!]
\begin{center}
  \includegraphics[width=1.00\textwidth]{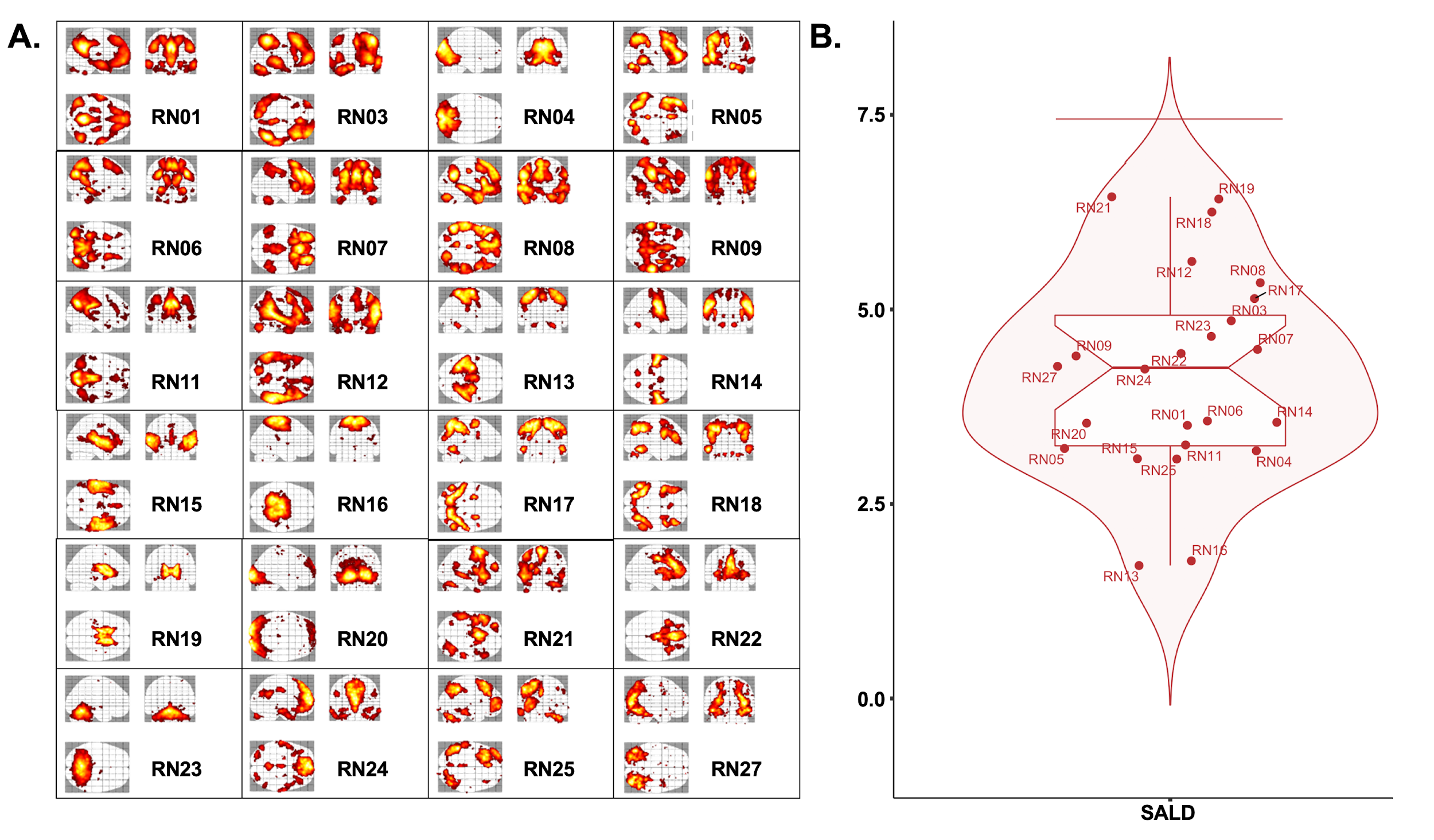}
\caption{Presentation of the Resting-state Networks ({\tt RNs}) of the {\tt SALD} cohort and their stability. \textbf{A.} Spatial map of the {\tt RNs} obtained from the multi-step independent component analysis on the intrinsic connectivity data of the {\tt SALD} cohort. \textbf{B.} Tukey box of the stability $Q_k$ score for each network in the {\tt SALD} cohort. The horizontal red line represents the limit at which a {\tt RN} is considered unstable (extremum).}
\label{fig:5}
\end{center}
\end{figure}

An example of the application of the method may be found in \cite{RefK}. 
There is currently no brain atlas of the intrinsic organization of the brain constructed from populations whose individuals are over 40 years of age. However, the brain, and thus the brain networks that compose it, undergo continuous reconfiguration throughout adult life (\cite{RefL}, \cite{RefM}). The absence of brain atlases from older populations therefore directly influences the results related to the different properties of intrinsic networks. In this context, the objective of the study was to construct a reliable brain atlas derived only from healthy older participants.
Doucet et al. \cite{RefK} analyzed resting-state fMRI data from $184$ individuals aged $55$-$80$ years from the {\tt SALD} cohort (Southwest University Lifespan Dataset, \cite{RefN}). Using a multi-step independent component analysis approach they identified $24$ Resting-state Networks ({\tt RNs}, Fig.~\ref{fig:5} - A.).

In this framework, our method was used to verify: 1) the stability of the identified intrinsic networks ({\tt RNs}) and 2) the presence of subpopulations in the data (the cophenetic threshold was fixed at  $0.85$). 

We found that no {\tt RNs} in the {\tt SALD} sample of individuals aged $55$ or older were unstable (Fig.~\ref{fig:5} - B.). 

\begin{figure}[htbp!]
\begin{center}
  \includegraphics[width=1.00\textwidth]{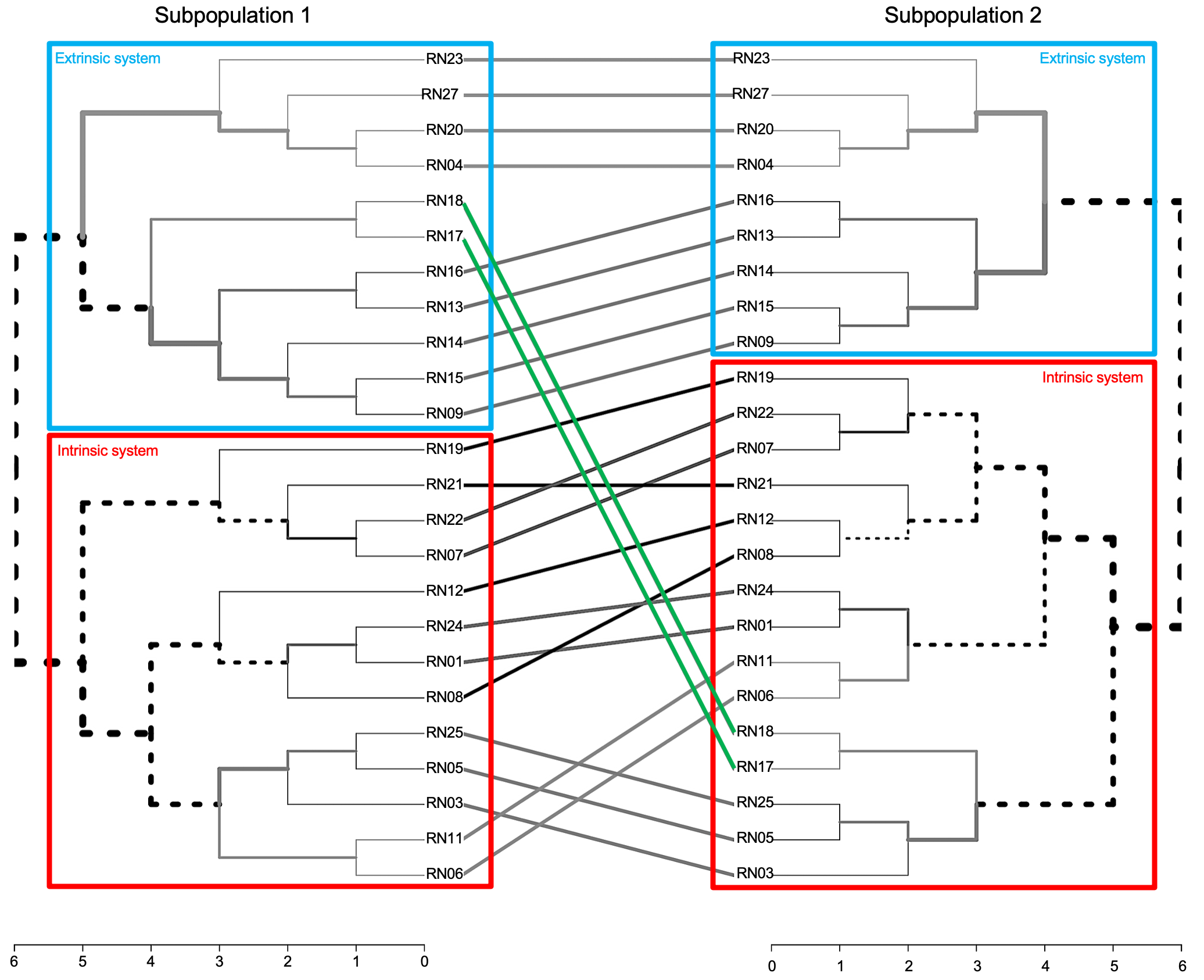}
\caption{Tanglegram between subpopulations in the {\tt SALD} cohort. The red rectangle represents the intrinsic system; in blue the extrinsic system. The green lines correspond to the two {\tt RNs} whose positioning in the mean dendrogram differs from one subpopulation to another.}
\label{fig:6}
\end{center}
\end{figure}

In addition, two subpopulations were identified in the {\tt SALD} cohort Fig.~\ref{fig:6}. The first one (Subpopulation $1$ in Fig.~\ref{fig:6}) is composed of 85 individuals and has a mean age of $62.5$ years old. The second one (Subpopulation $2$ in Fig.~\ref{fig:6}) is composed of $94$ individuals and has a mean average age of $64.5$ years old. A two-tailed Student’s t-test revealed no difference between the $2$ subpopulations in terms of demographics variables (Edinburgh test: $p = 0.8$, sexe: $p = 0.15$, Pearson's chi-squared test), neither in anatomical variables (Total Intracranial Volume: $p = 0.81$) or in quality control variable (all $p > 0.2$). A one-tailed t-test revealed a significant inferior average age in the first subpopulation than in second one ($p = 0.032$). 

A comparison of the average dendrograms of the two subpopulations highlights the existence of $2$ partitions in each subpopulation. The first one includes {\tt RNs} related to the extrinsic system (blue rectangles in Fig.~\ref{fig:6}): which is a system driven by external inputs and activated during sensory stimulations, including the attentional and the sensory-motor network. The second partition corresponds to the intrinsic system (Red rectangle in Fig.~\ref{fig:6}) that is used in inner-oriented mental activity. 

This organization of resting-state brain processes corresponds well to the organization highlighted in \cite{RefF}. The calculation of the cophenetic distance between the two means dendrograms of the two subpopulations was 0.76, revealing two slightly different organizations. The main difference between the two subpopulations (Fig.~\ref{fig:6} comes from the networks {\tt RN18} and {\tt RN17}, which in the case of subpopulation $1$, belongs to the extrinsic system whereas these {\tt RNs} belong to the intrinsic system in the second. The {\tt RN18} corresponds to one of the fronto-temporo-parietal network involved during the executive processes of selecting and monitoring our behaviours. The {\tt RN17} corresponds to the spatial attention network. These two {\tt RNs} are grouped together very early in the two dendrograms, thus we can tentatively  identify this couple as being a marker of the resting state activity  either turned towards the outside, or towards oneself. 

This organizational difference, possibly related to age, compared to young subjects \cite{RefF}, deserves to be addressed more in-depth study with complementary neuroscientific approach, such as the resting state questionnaire for evaluation of inner experience during the conscious resting state \cite{RefO}.

\section{Concluding remarks}
\label{conclu}
In this study, we introduced two statistical approches allowing: 1) to analyze the stability of the constituent features of an average dendrogram produced from three-dimensional data, which is the $Q_k$ score, and 2) to extract homogeneous subpopulations in terms of individual dendrograms from three-dimensional data. 

The first one is based on the new reliability $Q_k$ score introduced in Section~\ref{sec:2}. This core represents the relative position of the network k across the individual dendrograms, relative to the average dendrogram. The simulation study provided good numerical performance of the proposed methodology. 

Furthermore, the second proposed approach is able to identify the existence of a subpopulation of subjects with a pattern of features that differs constantly within it, or the existence of a subpopulation that does not have a strong structure in terms of dendrogram.

Concerning the capacities of the algorithm to identify subpopulations, the similarity between subpopulations, the proportion of each subpopulation directly influences the performance of the algorithm. A subpopulation which is small in number of individuals and whose mean dendrogram is not singular: i.e. whose cophenetic distance is high in relation to the mean dendrogram of the other subpopulations, will be difficult to identify. In contrast, a large subpopulation with a singular dendrogram will be easy to identify. The modulation of the threshold value t comes into play here, a t value close to 1 allowing a better distinction between two similar dendrograms according to the cophenetic distance.

These two new methods will soon be available via a library called {\tt SIMS} (\textit{Similarity of Individual MatriceS}) on {\tt GitHub}, on {\tt CRAN}, and it can also be directly requested from the first author.

\section*{Funding}
Gaelle E. Doucet was supported by the National Institute of Aging (R03AG064001) and the National Institute of General Medical Sciences (P20GM130447).

\section*{Conflict of interest}
The authors declare that they have no conflict of interest.

\bibliographystyle{spbasic}      


\end{document}